\documentclass[aps,article,twocolumn,superscriptaddress]{revtex4}
\usepackage{graphics} 
\usepackage{amsmath}
\usepackage{color}

\begin{document}

\title{Electron hole instability in linearly sub-critical plasmas}
\author{Debraj Mandal}
\author{Devendra Sharma}
\affiliation{Institute for Plasma Research, HBNI, Bhat, Gandhinagar, India, 382428}
\author{Hans Schamel}
\affiliation{Physikalisches Institut, Universit\"{a}t Bayreuth, D-95440 Bayreuth, Germany}

\date{\today}

\begin{abstract}
Electron holes (EH) are highly stable nonlinear structures met omnipresently 
in driven collisionless hot plasmas. A mechanism destabilizing small 
perturbations into holes is essential for an often witnessed but less 
understood subcritically driven intermittent plasma turbulence. In this 
paper we show how a tiny, eddy-like, non-topological seed fluctuation 
can trigger an unstable evolution deep in the linearly damped region, 
a process being controlled by the trapping nonlinearity and hence being 
beyond the realm of the Landau scenario. After a (transient) 
transition phase modes 
of the privileged spectrum of cnoidal EH are excited which 
in the present case consist of a solitary electron hole (SEH), two 
counter-propagating ``Langmuir'' modes (plasma oscillation), and an 
ion acoustic mode. 
A quantitative explanation involves 
employing nonlinear eigenmodes, yielding 
a nonlinear dispersion relation with a forbidden regime and the negative 
energy character of the SEH, properties being inherent in Schamel's model 
of undamped Vlasov-Poisson structures identified here as lowest order 
trapped particle equilibria. An 
important role in the final 
adaption of nonlinear plasma eigenmodes is played by 
a deterministic response of trapped electrons
which facilitates transfer of energy from electron thermal energy to
an ion acoustic nonuniformity,
accelerating the SEH and positioning it into the right 
place assigned by the theory.
\end{abstract}



\maketitle
Subcritically driven turbulence of plasma state remains a less understood
process, often presenting its strong signatures in nature 
\cite{petkaki:p, osmane:a}, experiments \cite{saeki:k,fox:w,colestock:p} 
and in simulations 
\cite{berma:r82,lesure:m09,berk:h,lesure:m,eliasson:b,eliasson:b06} 
of collisionless hot plasmas.
Underlying this are instabilities of nonlinear collective eigenmodes 
of nonthermal distributions rather than those of the normal linear 
eigenmodes of a thermalized 
distribution $f_{0}$, 
recoverable by selecting the
corresponding poles of dispersion function to perform the Landau integral,
yielding $f_{0}'\equiv \partial f_{0}/\partial v$ as a unique driver for the 
microinstabilites.
Explanation of this stronger nonlinear basis of the turbulence threshold 
is explored both by stochastic \cite{lesure:m,lesure:m1} as well as 
deterministic approaches \cite{dupree83:t,hutchinson:i}, prescribing the 
growth largely linked to species' $f'$. 
With these criteria often defied by the evolution, 
no 
basis is 
known for quantitatively exploring drivers of rather complex unstable 
subcritical evolution 
\cite{schamel:hms} of coherent phase-space 
perturbations 
constituting 
fundamental nonlinear collective eigenmodes in hot 
nonthermal collisionless plasma \cite{hutchinson:i17},
inevitably unstable if they possessed a forbidden regime or violated the 
negative energy state condition \cite{griessmeier:j} in certain regimes.
By first recovery of these two characteristic EH attributes in our 
simulations,
we have quantitatively applied, to the observed evolution, a formulation 
implementing a stochastic scale cut-off to approach fundamental smallest 
nonlinear unit of phase-space perturbations \cite{schamel:h1}. 
We have thus characterized the subcritcally unstable
response in terms of parameters that allow generalization to 
ensembles, or large scale nonthermal phase-space equilibria.

We present results of two cases of high-resolution Vlasov simulations 
initialized with small phase-space perturbations capable of developing 
into unstable hole 
structures. A forbidden regime is identified for the electron holes where
they accelerate providing 
evidence of their multifaceted 
sub-critical nonlinear instability \cite{schamel:hms, holloway:j}, growing 
coherent structures. The second part of observations shows that the 
electron holes can also be destabilized by parametric coupling to 
conventional collective modes of collisionless plasmas. 
In all cases the phase velocity $v_{0}$ of the finally settled SEH exceeds 
the electron drift and is hence located at the right wing of $f_{e0}$, 
which has a negative 
slope that, according to standard wave theory, would imply disappearance 
by Landau damping \cite{landau:ld1946}. We hence have observed a nonlinear evolution
beyond the generally accepted Landau scenario for the plasma turbulence.

For the present exact mass ratio simulations ($\delta=m_e/m_i = 1/1836$) 
we have used 
a well localized initial perturbation in the electron 
distribution function of the following analytic form,
\begin{eqnarray}
f_{1}(x,v)= - \epsilon~{\rm sech}\left[\frac{v-v_1}{L_{1}}\right]{\rm sech}^4[k(x-x_{1})] 
\label{perturbation}
\end{eqnarray}
where $\epsilon$ is the amplitude of the perturbation, $L_{1}$ is the
width of the perturbation in the velocity dimension and $k^{-1}$ is its 
spatial width. 
We use the Debye length $\lambda_{D}$, electron plasma frequency 
$\omega_{pe}$ and electron thermal velocity $v_{the}=\sqrt{{T_e}/{m_e}}$ 
as normalizations 
for length, time and electron velocities, respectively.  
The background electron and ion velocity distributions are
Maxwellian with a finite electron drift $v_D$, 
\begin{eqnarray}
f_{0e}(v)&=&\frac{1}{\sqrt{2 \pi}}\exp\left[-\frac{(v-v_D)^2}{2}\right]\\
f_{0i}(u)&=&\frac{1}{\sqrt{2 \pi}}\exp\left[-\frac{u^2}{2}\right]
\label{initial_f}
\end{eqnarray}
where 
$u=\sqrt{\theta/\delta}~v$ and
$v_D=0.01$ is chosen below the critical linear threshold $v^*_D=0.053$ 
\cite{fried:b} for
$\theta=T_e/T_i = 10$ used by us. 

We first present the evolution of the total electron distribution 
$f_{e}=f_{0e}+f_{1}$ 
in two cases, 1 and 2, where $v_{1}=0.05$ and $0.01$, respectively,
i.e., the perturbation located beyond the maximum of 
$f_{0e}(v)$ far in the decreasing tail in case 1 and
just at its maximum in case 2.
It is additionally initiated from the center, 
$x=15 $, of the simulation box of length $L=30$. 
The phase-space
widths of the perturbation is chosen as $L_{1}=0.01$ along
$v$ and $k^{-1}=10$ along 
$x$ in expression (\ref{perturbation}) with
the perturbation strength $\epsilon=0.06$. 

\begin{figure}
\includegraphics{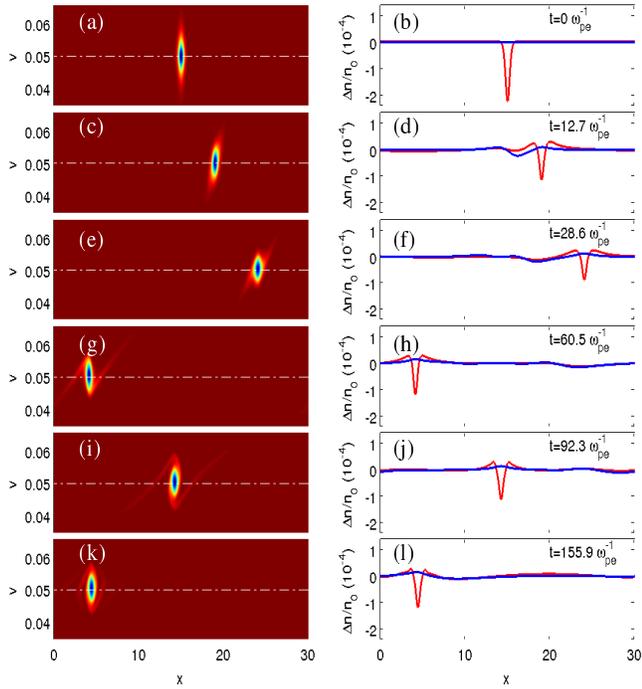}
\caption{Evolution of the electron phase-space perturbation and the density 
perturbations initially introduced at $(x_{1},v_{1})=(15,0.05)$.
\label{case1}}
\end{figure}

\begin{figure}
\includegraphics{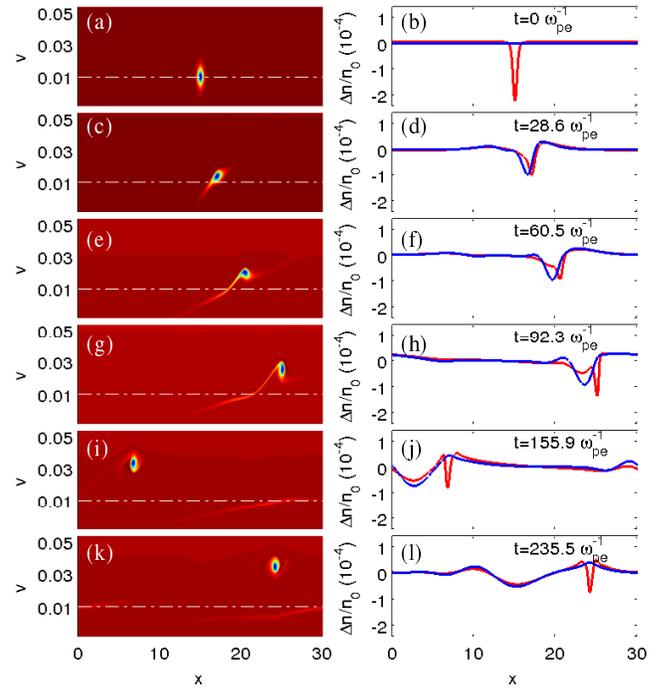}
\caption{Evolution of the electron phase-space perturbation and density 
perturbations initially introduced at $(x_{1},v_{1})=(15,0.01)$.
\label{case2}}
\end{figure}

Despite the fact that we are well in the linearly Landau damped region we 
expect the nonlinear excitation of an electron hole mode (EH), as suggested 
by our previous publication \cite{schamel:hms}. 
This EH is indeed recovered (Fig.~\ref{case1}) in case 1 
(apart from a completely decoupled undamped electron plasma oscillation 
in both cases)
where a much faster saturation of ion expulsion (potential decay) is 
achieved resulting in an immediate set up of a coherently propagating 
structure. 
For the second slower perturbation, however, the 
phase-space structure presented in left column of Fig.~\ref{case2} is 
seen accelerating to a higher velocity after a noticeable change in its 
topology in the phase-space.
%
For both these cases, 
the 
removal of electrons ($f_{1}\ll f_{0}$) from a small velocity interval 
translates in an electron density dip (potential hump) at $x_{1}$, 
instantly introducing a phase-space separatrix about ($x_{1},v_{1}$).
A slowly varying separatrix corresponds to an
adiabatic invariant, with a response time (time for it to modify) 
longer than that of untrapped ions ($\tau_{\rm adiabatic}\gg\omega_{ip}^{-1}$). 
While ions can be expelled faster to restore quasineutrality, an
inward flux of them is also expected, driven by deficiency of thermal
electrons at $x_{1}$ that must allow ions to easily bunch at $x_{1}$
\cite{anderegg:f}.
Clearly, in a stably propagating solitary electron-hole structure, these two 
fluxes must balance and a comoving ion density hump must exist, as seen in 
Fig.~\ref{case1}(j). 
However, an unstable, subcritically evolving and accelerating perturbation
recovered in Fig.~\ref{case2}, in clear contrast to Fig.~\ref{case1}, 
is subject of this letter.

Note that 
in both cases the $u_1$'s are sufficiently large (6.78 and 1.36 $v_{thi}$),
to neglect ion trapping in first approximation. 
However, 
since the ion sound speed $c_s=3.16 v_{thi}$, 
in case 1 the perturbation is moving 
supersonically, 
in case 2 we have a subsonic propagation. This implies that 
the ion mobility can be largely neglected in case 1 but plays an important 
role in case 2. Consequently, the time scales of the evolution are rather 
distinct in both cases, being determined essentially by $\omega_{pe}^{-1}$ 
for case 1 where the EH has settled in about 10 $\omega_{pe}^{-1}$,  
but by $\omega_{pi}^{-1}$ for case 2 where the settling occurs in about 
3.6 $ \omega_{pi}^{-1} \approx 156 \omega_{pe}^{-1} $.
The present simulations therefore indicates a gap of existence for the 
electron hole solutions which is now addressed for the first time 
well within 
the analytic model for equilibrium 
solutions of the Vlasov-Poisson system presented by Schamel where 
the trapped particle effects are retained in distributions. 

With extendability of the fundamental model of trapped species distribution 
$f_{st}$ to more deterministic forms (discussed further below),
the distributions $f_{i,e}$ are written by Schamel as function of 
total energy $\varepsilon_{i.e}$, hence satisfying the Vlasov equation
(see \cite{schamel:h1} and references therein). 
Using them
in Poisson's equation 
one can derive the nonlinear dispersion relation (NDR) (see equation (24) of 
\cite{schamel:h2}),
\begin{eqnarray}\nonumber
k_{0}^2&-&\frac{1}{2}Z_{r}^{'}\left(\tilde{v}_{D}/\sqrt{2}\right)
-\frac{\theta}{2}Z_{r}^{'}\left(u_{0}/\sqrt{2}\right)\\
&=&\frac{16}{15}\left[\frac{3}{2}b(\alpha,u_{0})\theta^{3/2}+
b(\beta,\tilde{v}_{D})\right]\psi^{1/2} ,
\label{NDR_eq}
\end{eqnarray}
where $Z_{r}(x)$ is the real part of the plasma dispersion function, 
$\tilde{v}_{D}:= v_{D}-v_{0}$ and $v_{D}$ 
describes a given constant drift between electron and ion existing already 
in unperturbed state.
The quantities $b(\alpha,u_{0})$ and $b(\beta,\tilde{v}_{D})$ are given by,
\begin{eqnarray}\nonumber
b(\alpha,u_{0})=\frac{1}{\sqrt{\pi}}\left(1-\alpha-u_{0}^{2}\right)
\exp(-u_{0}^{2}/2) \\
b(\beta,\tilde{v}_{D})=\frac{1}{\sqrt{\pi}}\left(1-\beta-\tilde{v}_{D}^{2}
\right)
\exp(-\tilde{v}_{D}^2/2)\nonumber
\label{b_beta}
\end{eqnarray}
respectively, where $\beta$ and $\alpha$ are the trapping parameters for 
electrons and ions with $b(\alpha,u_{0})=0$ for no trapping effects of ions. 
The NDR (\ref{NDR_eq}) determines the phase velocity of structures ($v_{0}$
or $u_{0}$) in terms of $v_{D}$, $k_{0}^2$, $\theta$, $\psi$, 
$\alpha$ and $\beta$.
The corresponding pseudo-potential $V(\phi)$ in case of no ion trapping is 
given 
(see (25) of  \cite{schamel:h2}) by:
\begin{eqnarray}\nonumber
-V(\phi)= \frac{k_0^2}{2} \phi (\psi - \phi) + \frac{B}{2} \phi^2(1 - \sqrt{\phi/\psi}), 
\end{eqnarray} 
where 
\begin{eqnarray}\nonumber
B:= \frac{16}{15} b(\beta,\tilde{v}_{D}) \sqrt \psi.
\end{eqnarray} 

In generality, we  meet a two parametric solution (described by the 
parameters $k_0^2$ and $B$), 
which is termed cnoidal electron hole (CEH) because it can be expressed 
by Jacobian elliptic functions such as $cn(x)$ or $sn(x)$. 
It incorporates as special cases the familiar solitary electron hole (SEH), 
when $k_{0}^{2}=0$ and $B>0$ \cite{schamel:h4,schamel:h5}, the harmonic wave, 
when $B=0$, as well as the special solitary potential dip (SPD), when 
$k_{0}^2= -\frac{4B}{2}>0$ demanding $B<0$.
\begin{figure}[t]
\includegraphics{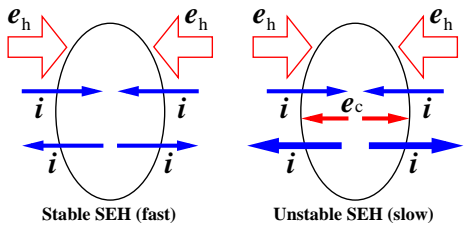}
\caption{Schematic of (left) valid fast SEH with $v_{0}\geq 0.028$ (right) unstable
slow SEH in the forbidden regime $v_{0}< 0.028$.
\label{schematic}}
\end{figure}

{\em Saturated holes as valid SEH solutions} : 
We now validate holes settled in equilibrium states as described by above
analytic model.
Since our code is periodic the lowest available wavenumber 
is $k_0=\frac{2\pi}{L}=0.21$, ($k_0^2=0.04$), to
approximate SEH with. We moreover recognize that both 
$v_D$ and 
$v_0$, and hence $\tilde v_D$, are 
small quantities such that $-\frac{1}{2}Z'_r(\tilde v_D / \sqrt 2) \approx 1$ 
to a good approximation, while noticing that $Z'_r(x)$ is an even 
function. 
Under these special conditions our NDR simplifies and 
becomes, in case of negligible ion trapping:
\begin{eqnarray}
-\frac{1}{2}Z'_r(u_0 / \sqrt 2)=\frac{1}{\theta}[B-(1+k_0^2)] \equiv 
\frac{B - 1.04}{10}=:D
\label{NDR_simplified}
\end{eqnarray}
An inspection of the  $-\frac{1}{2}Z'_r(x)$ shows (see Fig.~1 of 
\cite{schamel:h1}) that $D$ is negative, corresponding to $1.307 < u_0$, 
provided that $0<B<1.04$. Taking the ideal SEH solution, 
$\phi(x)=\psi {\rm sech}^4(\frac{x}{\Delta})$  with $\Delta=\frac{4}{\sqrt B}$, 
this amounts to $\Delta > 3.92$. Since the spatial width of our perturbation 
is essentially maintained during the evolution we can take the initial width 
and approximate $\Delta$ by $\Delta \approx \frac{1}{k} = 10$ such that $B$ 
becomes $B\approx 0.16$. On the other hand, $B$ is given in the present 
situation by $B=\frac{16(1-\beta)\sqrt \psi}{15 \sqrt \pi}$, which gives, 
for $\psi \approx 10^{-4}$, a value of the electron trapping parameter 
$\beta \approx -25.6$. Analytically, we hence get a depression of the 
electron distribution in the resonant or trapping region, as observed. 
The corresponding phase velocity is for this case with $D \approx -0.09$ is 
found to be $u_0 \approx 3.7$ or $v_0 \approx 0.027$, i.e. in the observed 
range.

{\em The acceleration of SEH}: 
The function 
$-\frac{1}{2}Z'_r(x )$ has a minimum of  $-0.285$ at $x=1.5$, which 
corresponds in terms of $u_0$ in (\ref{NDR_simplified}) to $u_0=2.12$. 
This yields, by 
use of 
(\ref{NDR_simplified}), $B=-1.81$ which is outside the admissible 
range of $B$, $0<B<1.04$. There is hence a gap in $u_0$ in which no 
equilibrium (quasi) SEH can exist. 
The lowest value of $B$ for which a solution exist is $B=0^+$ corresponding 
to $D=-0.104$ or $u_{0s}=1.48$  ($x_s=1.05$) and $u_{0f}=3.61$ ($x_f=2.55$),
hence a gap bounded by these slow and fast velocities,
$1.48< u_0 < 3.61$.
This explains why a slow perturbation in case 2 
($v_1=0.01\equiv u_1=1.36$), which despite acquiring an adiabatic character, 
cannot settle below $u_0=3.61$.
The simulations with much slower perturbation $v_{1}=0.004$ (not presented 
here) additionally showed that the acceleration continues despite the condition 
$f'_{i}f'_{e}<0$ 
\cite{dupree83:t}
was violated when EH velocity crossed $v_{D}$. 
It remains to be shown as to why the hole must accelerate, instead of 
decaying by phase mixing or decelerating. Quantitatively supported by 
the energy balance presented further below, the mechanism underlying 
this acceleration is well explained by the simulated phase-space evolution 
of the hole, illustrated more clearly in the schematic Fig.~\ref{schematic}. 
While the net charge flux is balanced (zero)
for the fast moving structures (left), in a slow moving structure (right) 
the inbound ion flux limited by finite $T_{e}$ is too weak to balance the 
outbound ion flux generated by a longer exposure to
hole electric field, 
$\Delta t \sim 4\pi/v_{0}k$. With finite trapped electron 
population, 
this insufficient ion influx in a slow moving hole is supplemented by 
the deterministic response of
trapped electrons which create an effective flux by beginning to update 
their phase-space orbits.
The spatial distribution 
of trapped electrons keeps modifying until the saturation, effectively 
increasing $|\beta|$, and hence increasing the hole velocity \cite{schamel:h5}.
Note that interpreting $\beta^{-1}$ as trapped electron 
temperature (i.e. $f_{et}$ a maximum entropy state), 
lets the EH represent an structure of infinitesimal scale below which no 
internal phase-space structures are considered. 
For treating a deterministic (Vlasov) prescription of internally 
structured finite amplitude EH, this opens possibility of 
generalizing Schamel approach by defining a multitude of $f_{jt}$s, $j=e,i$, 
(in mutual equilibrium, e.g., in phase locked states \cite{dodin:i,smith:g}) 
with an associated set of $\beta$s and $\alpha$s.

{\em Parametric phase of EH instability}: 
Additionally seen in our results is a further acceleration 
continuing beyond
$t=92.3$ ((g),(h) in Fig.~\ref{case2} where $v_0= 0.028$ or $u_0=3.8$) when $B$ 
changes sign to become positive.
The further increase in 
$v_0$ ($u_0$) at later times is due to an increase of $D$ (decrease of $|D|$) 
or increase of $B=\frac{16(1-\beta)\sqrt \psi}{15 \sqrt \pi}$. The latter can 
have two sources, an increase of $\psi$ and 
an increase of  
$(1-\beta)=(1+|\beta|)$, corresponding to a deeper (or sharper, with large $k$) 
depression in the phase space vortex center. 
This additional acceleration essentially corresponds to a net imbalance of 
ion flux across the separatrix of a valid hole ($B>0$) due to 
difference in ion density
at two ends of the hole, or an ambient ion density 
gradient, that must cause further trapped electron response, 
and hence the acceleration. 
The model \cite{schamel:h2} can hence explain both, a gap in $u_0$ and a further acceleration 
along the fast dispersion branch.

{\em Negative energy character of settled hole}: 
As derived in \cite{griessmeier:j,griessmeier:j2,luque:a} the total 
energy density $w$ of a SEH carrying plasma is changed by
\begin{eqnarray}
\Delta w =\frac{\psi}{2}[1 + \frac{1}{2}Z'_r\left(\frac{u_0}{\sqrt 2}\right)(1-u_0^2)]
\label{negative-energy}
\end{eqnarray}
with respect to the unperturbed, homogeneous state. This expression is 
negative  when it holds: $2.12 < u_0$ (see Fig.~1 of \cite{griessmeier:j}), 
and is satisfied for all of our settled SEHs. 
\begin{figure}[t]
\includegraphics{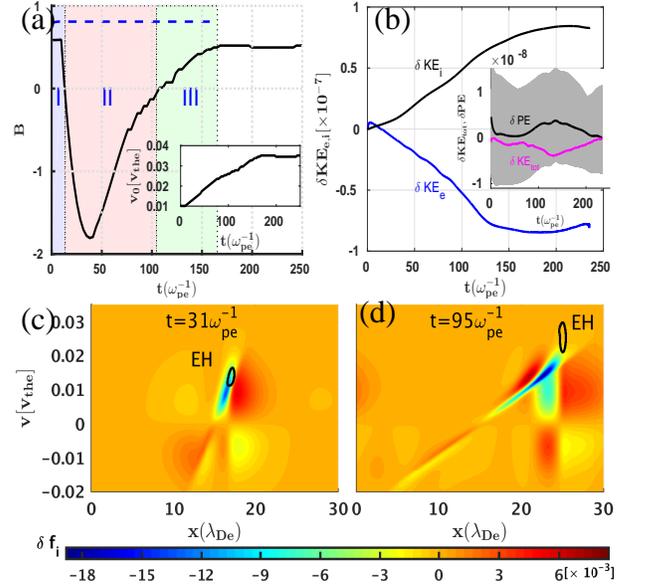}
\caption{(a) Time evolution of parameter $B$ and (subplot) velocity 
$v_{0}$ of the SEH. (b) Time variation of change from initial value of
kinetic energy of ions (black line), and electrons (blue, averaged over 
fast electron oscillations). 
The subplot shows variation of potential (gray and black, total and 
averaged, respectively) and total kinetic energy (magenta). 
(c) and (d) $\delta f_{i}$ in the ion phase-space at indicated
times and black solid line is contour of $f_{e}$ representing the 
electron hole.
\label{energy_conserv}}
\end{figure}

{\em Time evolution of trapped species dispersion $B$}: 
In Fig.~\ref{energy_conserv}(a) we have presented the 
time variation of $B$, calculated using Eq.~(\ref{NDR_simplified}) 
and rest of the quantities available from the simulation data.
This can be noted that for the case 1 (represented by the dashed line) 
$B$ is uniform and positive at all times as required for the valid SEH 
eigenmode. For the case 2, however, the value of $B$ (solid line) has
negative value in a finite interval (region II) indicating no valid SEH 
eigenmodes, explaining the unshielded phase of the SEH during this 
initial interval in case 2. 
The initial $B>0$ phase  ($t<10$ or region I) in this case still has an 
inappropriate SEH eigenmode that violates the negative energy state 
condition $v_{0}\geq 0.016$ as described by 
(\ref{negative-energy}) \cite{griessmeier:j}.
The value of $B$ can still be seen changing once
the unshielded phase (region II) is over and $B>0$ is achieved 
which is attributed to interaction of valid SEH with the background ion 
acoustic structure created during the unshielded phase. This is evident 
from the saturation in the $B$ variation that exactly corresponds to the 
time of exit of the SEH from the region of a positive ion density gradient 
($t\sim 153 \omega_{pe}^{-1}$ in Fig.~\ref{case2} and \ref{energy_conserv}(a)).

We now show that the energy to accelerate ions and growth of an ion acoustic 
perturbation is derived from the thermal energy of electrons,
establishing the unstable hole evolution as a fundamental mechanism for 
the plasma destabilization driven by a source of free energy.
The resulting ion acoustic structure in turn interacts with the SEH and 
accelerates it further in $B>0$ regime. 
This exchange is mediated by the deterministically modifying trapped 
electron phase-space orbits that allow the electron hole to 
survive the otherwise expected steady decay of its electric field caused 
by its unshielded phase.

The steady growth in the ion kinetic energy ($\delta$KE$_{i}$) and a
corresponding loss of the averaged thermal energy of the streaming (hot) 
electrons 
($\delta$KE$_{e}$), are plotted in Fig.~\ref{energy_conserv}(b) indicating 
the conversion of $\delta$KE$_{e}$ into $\delta$KE$_{i}$ \cite{luque:a}. 
This energy 
exchange is mediated by trapped electrons whose trajectories modify with 
time, allowing them spend longer time away-from/close-to center 
in unstable $B<0$ regime, to supplement
the incoming ion flux 
(pushed by excess thermal electrons) that would balance the outgoing ion 
flux in a valid plasma eigenmode. 
The higher $|\beta|$ values correspond
to larger dispelled density of trapped electrons and, in turn, to
higher hole velocity, explaining the hole acceleration for $t<\tau$,
that continues in region III due to coupling with ion density 
nonuniformity.
This conversion of electron thermal energy to ion kinetic energy however 
need not be 100 \% as a fraction of variation in the total thermal energy 
($\delta {\rm KE}_{\rm tot}=\delta {\rm KE}_{e}+\delta KE_{i}$)
balances that in the sum ($\delta$PE) of electrostatic energies of the SEH and 
the developing 
ion compression wave structures
(plotted for case 2 in the subplot of Fig.~\ref{energy_conserv}(b) as 
magenta line and black line, respectively).
We have also presented the entire process in the ion phase-space by plotting
$\delta f_{i}=f_{i}-f_{i0}$ at two time points in 
Fig.~\ref{energy_conserv}(c) and (d). The contour of SEH separatrix 
is superimposed at both the times on the contours of $\delta f_{i}$
where an SEH with $B<0 (t<\tau)$ can be seen coinciding large 
$\partial \delta f_{i}/\partial x$, while a valid SEH, with trapped 
electrons coinciding the ion density hump is seen for $B>0 (t>\tau)$.

To summarize, we have indicated presence of a new forbidden regime of nonlinear 
electron hole structures at smaller velocities in linearly subcritical
collisionless plasmas. 
The evolution is shown to be manifestation of an already predicted 
\cite{schamel:hms} multifaceted nonlinear 
SEH instability modifying parameters other than the 
structure amplitude available for the linear eigenmodes. 
Importantly, the independence of the nonlinear evolution from the
$f'$ and the role of trapped particles that 
facilitate conversion of thermal energy to coherent modes show the 
observed evolution beyond the realm of linear Landau scenario.
Finally we mention that our analysis rests on the availability of a NDR, 
which is provided by the used method treating a basic nonlinear 
eigenmode. A BGK-analysis, 
in its purity,
could not be applied 
because of the lack of a NDR, which is a consequence of the 
strong slope singularity of the derived $f_{et}$  within the BGK method 
\cite{SDB18}.

By establishing negative energy SEHs the plasma gains free energy and 
resides in a metastable, structural thermodynamic state. In the long term 
run, when dissipative processes are no longer negligible, this enables the 
plasma to heat electrons and approach the thermodynamic equilibrium state 
faster than without 
this intermediate structural state. The latter property is suggested by the 
existence of separatrices around which collisionality is appreciably 
enhanced.

\end{document}